\documentclass[preprint,12pt,3p,superscriptaddress]{elsarticle}

\usepackage{amssymb}
\usepackage{fullpage}
\usepackage{times}
\usepackage{subfigure}
\usepackage{fancyheadings}
\usepackage[cp1251]{inputenc}
\usepackage[english]{babel}
\usepackage{graphicx}
\usepackage{color}
\usepackage{amsmath,amssymb,amsthm,enumerate,bbm,color,verbatim,setspace}

\newcommand{\Ignore}[1]{ }

\begin{document}

\begin{frontmatter}

\title{An example of interplay between Physics and Mathematics:\\ Exact resolution of a new class of Riccati Equations}
%\tnotetext[label0]{This is only an example}

\author{L.A. Markovich}
\address{V.A. Trapeznikov Institute of Control Sciences, Moscow, Profsoyuznaya 65, 117997 Moscow, Russia.}
\address{Institute for information transmission problems, Moscow, Bolshoy Karetny per. 19, build.1, Moscow 127051, Russia.}
\address{Moscow Institute of Physics and Technology, Institutskii Per. 9, Dolgoprudny Moscow Region 141700, Russia.}

\author{R. Grimaudo, A. Messina}
\address{Dipartimento di Fisica e Chimica dell'Universit\`a di Palermo, Via Archirafi, 36, I-90123 Palermo, Italy.}
\address{I.N.F.N., Sezione di Catania, Catania, Italy.}

\author{H. Nakazato}
\address{Department of Physics, Waseda University, Tokyo 169-8555, Japan.}

\Ignore{
\author{$^{1,2,3*}$, $^{4}$,  A. Messina$^{4,5}$, H. Nakazato$^{6}$\\
$^1$\\
$^2$\\
$^3$\\
$^4$\\
$^5$Sezione I.N.F.N., Catania\\
$^6$\\
$^*$Corresponding author e-mail: kimo1@mail.ru}
}
\date{}

\begin{abstract}
A novel recipe for exactly solving in finite terms a class of special differential Riccati equations is reported.
Our procedure is entirely based on a successful resolution strategy quite recently applied to quantum dynamical time-dependent SU(2) problems.
The general integral of exemplary differential Riccati equations, not previously considered in the specialized literature, is explicitly determined to illustrate both mathematical usefulness and easiness of applicability of our proposed treatment.
The possibility of exploiting the general integral of a given differential Riccati equation to solve an SU(2) quantum dynamical problem, is succinctly pointed out.
\end{abstract}

\begin{keyword}
Riccati equation, solvable model, time-dependent Hamiltonian, unitary evolution
\end{keyword}

\end{frontmatter}

\section{Introduction}
Theoretically investigating the properties of systems in different contexts \cite{Reid,Fraga,Laub}, such as for example classical and quantum physics \cite{Schuch,Schuch1,Rau,Rau1,Harko,Liverts,Unal,Gardas,Bastami,Suazo,Kus,Kus1,Kus2}, mathematics \cite{Bougoffa,Busawon,Messina,Mukherjee,jivulescu,Panayotounakos,Tesavrova,Yamaleev,Yuzbasi}, biology \cite{Kengne,Navickas}, one is led to the consideration of the following non-linear non-autonomous first order differential equation
\begin{equation}\label{DRE}
y'(x) = A(x)y^2(x)+B(x)y(x)+C(x).
\end{equation}
Here $y$ and $x$ are the dependent and independent variables, respectively, and $y'(x)$ denotes the first derivative of $y$ with respect to $x$.
The intrinsic nature of the variables $x$ and $y$ depends of course on the problem under scrutiny.
Such an equation is known as the scalar Differential Riccati Equation (DRE) and the three coefficient functions are supposed to be continuous in some interval $I \in \mathbb{R}$ \citep{Bittanti,Polyanin,Zwillinger}.
To emphasize how wide is the range of different situations where such an equation emerges, it has been suggested that the DRE, as given by Eq. \eqref{DRE}, should be included in the list of the named equations of mathematical physics (\citep{Hasegawa} and references therein).

One should not be too surprised by such a central role played by the DRE.
Indeed, in 1769 (\citep{Bittanti} and references therein), Euler demonstrated that every given DRE may be always transformed into a second-order linear, generally non-autonomous, differential equation and vice versa.
Thus, since such differential equations represent very often the starting point for theoretically investigating both classical and quantum systems \cite{Klimov,Marmo}, in accordance with first principles \cite{Haley,Ronveaux,Ronveaux1,Santiago,Kryachko} or as a result of an approximation path \cite{Hasegawa,Sidharth,Bougouffa,Filho}, it becomes easy to understand the systematic emergence of an associated DRE in the problems under scrutiny.

The circumstance that no method is available to construct a particular integral of a DRE and the fact that, as well, no general protocol exists leading to the unique explicit solution of a Cauchy problem relative to a linear second order non-autonomous differential equation (2nd-LDE), may be seen as two different faces of the same mathematical difficulty.
Then, the intimate link existing between the two classes of mathematical problems assures the possibility of writing down the general integrals of both the linear and the related non-linear differential equations as soon as one is able to exactly solve one of them.
Such an expectation has spurred the research reported in the present paper, where the knowledge of the exact quantum dynamics of some SU(2) problems is exploited to bring to light a new class of exactly solvable DRE.
Here ``new'' means that, to the best of the present authors' knowledge, DREs possessing the particular relationship between the three coefficients, we are going to evidence in the next sections, do not appear in the most commonly used handbooks \cite{Polyanin,Zwillinger} and apparently are not present in more recent literature as well.

Our starting point is the quantum dynamics of a semiclassical generalized Rabi system, here defined as a spin 1/2 subjected to an arbitrary time-dependent magnetic field.
To find the evolution operator $U$ related to an SU(2)-symmetry dynamics is equivalent to solve a Cauchy problem for a homogeneous system of two coupled first-order differential equations in the two time-dependent complex unknown entries of $U$, which, in addition, must satisfy the usual unitarity condition.
This mixed system may always be converted into a new Cauchy problem related to a second-order linear non-autonomous differential equation which, in turn, may be transformed into a Cauchy-Riccati problem.

Generally speaking, no procedure is known to find, in finite terms, the time-dependence of $U$ when no special prescription is postulated on the time-dependence of the magnetic field.
Quite recently, however, novel resolution strategies for exactly solving the quantum dynamics of the SU(2) problem under scrutiny has been reported \cite{Mess-Nak,Kuna,Bagrov,DasSarma}.
The approach in Ref. \cite{Mess-Nak}, in particular, succeeds in singling out classes of exactly solvable time-dependent SU(2) Hamiltonian models, providing at the same time the successful protocol leading to the explicit construction of the relative evolution operators $U$.

In this paper we show how to turn the knowledge of the solution of such a physical problem into a recipe to solve a mathematical problem of general interest.
Our procedure consists of two steps.
First, we introduce a simple transformation by which the Cauchy problem for the determination of the two entries of $U$ is turned into a Cauchy-Riccati problem.
In this way we arrive at a special class $E$ of DREs whose coefficients are directly linked to the three time-dependent components of the magnetic field acting upon the spin.
At this point the knowledge of a particular solution of the quantum dynamics of the spin 1/2 enables us to write down, systematically and in a simple way, the explicit solution of the associated Riccati-Cauchy problem.
Such a solution, being a particular solution of the DRE under scrutiny, allows in turn the construction \citep{Reid,Hasegawa} of the general integral of the DRE.
To bring to light new approaches leading to classes of special DREs, for which an exact solution can be found, may be considered a noticeable mathematical result \textit{per se}.
%According with the approach of Ref. \cite{Mess-Nak} our second step consists in singling out a wide subclass of $E$ of exactly solvable DREs, providing the explicit construction of a particular solution and then of the corresponding general integral.
%In words, in this paper we show how to take advantage from the knowledge of the solution of a physical problem to bring to light a recipe to solve a mathematical problem of general interest.
%The following question, thus, immediately arises: {\color{red}based on the expectation that such a protocol is successful too for building the general integral of the DRE associated to each individuated exactly treatable Hamiltonian model, does such an exactly solvable DRE represent new examples of exactly solvable DRE?}
%In the next sections we explicitly determine both the structure of the new DRE and its general integral.

The paper is organized as follows.
The approach of Ref. \cite{Mess-Nak} to deduce the class of solvable SU(2) Hamiltonian problems is briefly reviewed in Sec. \ref{sec:2}.
New special DRE and their general integrals are then derived in Sec. \ref{sec:3} on the basis of Ref. \cite{Mess-Nak}.
Illustrative examples and conclusive remarks are finally given in Secs. \ref{sec:4} and \ref{sec:5}, respectively.

\section{A Short Review of the Approach in Ref. \cite{Mess-Nak}}\label{sec:2}

Our physical system is a generalized Rabi system whose Hamiltonian $H$, assumed generally time-dependent, may be given as
\begin{eqnarray}\label{16_45}
					H(t)=\left(
                     \begin{array}{cc}
                       \Omega(t) & \omega(t) \\
                       \omega(t)^{*} & -\Omega(t) \\
                     \end{array}
                   \right),%\quad H=H^{\dagger},\quad TrH=0,
\end{eqnarray}
%where $\Omega(t)$ and  $\omega(t)$ are functions of time.
The unitary evolution operator $U(t)$ generated by the Hamiltonian is an element of the SU(2)-group
\begin{eqnarray}\label{16_20}\text{SU}(2)=\Bigg\{U(t)=\left(
                                                             \begin{array}{cc}
                                                               a(t)& b(t) \\
                                                            -b^{*}(t) & a^{*}(t) \\
                                                             \end{array}
                                                           \right),\quad |a(t)|^2+|b(t)|^2=1,\quad a(t),b(t)\in\mathbf{C}
\Bigg\}
\end{eqnarray}
and satisfies the following Schr\"odinger-Liouville-Cauchy problem %equation as well as the indicated initial condition
\begin{eqnarray}\label{16_1}i\dot{U}=HU,\quad U(0)=I.
\end{eqnarray}
Equation \eqref{16_1} %immediately leads to the following Cauchy problem
%\begin{equation}\label{16_32}
%\left\{
%\begin{aligned}
%&i(\dot{a}a^{*}+\dot{b}b^{*})=\Omega,\quad \\
%&i(a\dot{b}-\dot{a}b)=\omega,\
%&a(0)=1,\quad b(0)=0
%\end{aligned}\right.
%\end{equation}
%for the two unknown entries of $U$, which, {\color{blue}due to the initial conditions}, satisfy the equation  $|a|^2+|b|^2=1$ at any time instant $t$. This 
immediately leads to the following Cauchy problem for the two entries $a(t)$ and $b(t)$ of $U(t)$ (the time-dependence is omitted for easier reading):
  \begin{equation}\label{16_42}
  \left\{
  \begin{aligned}
  &\dot{a}=-i\Omega a+i\omega b^{*},\\
  &\dot{b}=-i\omega a^{*}-i\Omega b,\\
  &a(0)=1,\quad b(0)=0.
  \end{aligned}
  \right.
  \end{equation}

  In accordance with the traditional procedure of resolution of a linear system of differential equations of the first order, one seeks the second-order linear non-autonomous  differential equation
  in $a(t)$ or $b(t)$. Unfortunately,  the resulting equation even if linear, cannot be solved unless some special links among its variable coefficients
  are given. The approach reported in  \cite{Mess-Nak}, successfully reaches the objective of providing a useful strategy for ``constructing'' solvable
  SU(2) problems, togehther their exact solutions. The method consists in introducing an auxiliary function $X(t)$ enabling the explicit analytical
    representation of both $a(t)$ and $b(t)$ at the cost of precisely defining the specific Hamiltonian of the exactly solvable SU(2) problem within
    the same resolution protocol. In other words one may claim that $X(t)$
  %  \begin{eqnarray}\label{16_4_1}X(t)\equiv\int_{0}^{t}\frac{\omega(t')}{a^2(t')}dt',\quad |a|^2\left(1+\frac{1}{\hbar^2}|X|^2\right)=1,\quad X(0)=0.
%\end{eqnarray}
generates too the peculiar link between the time dependent components of the applied magnetic field
    from the consistent construction of part of the solution.
%     \begin{eqnarray}\label{16_26}b=\frac{a}{i\hbar}\int_{0}^{t}\frac{\omega(t')}{a^2(t')}dt',
%\end{eqnarray}
%\begin{eqnarray}\label{16_4}\dot{a}=-\frac{\omega}{\hbar^2}X^{*}a^{*}+\frac{\Omega}{i\hbar}a.
%\end{eqnarray}
It is worthy to emphasize that, generally speaking, if such a link were given at will, there would be
    no certainty of being able to exactly solve the corresponding dynamical problem, for example starting from equations \eqref{16_42}. Thus, the merit of the approach
    under scrutiny is just that of furnishing a self-consistent recipe to single out solvable SU(2) problems and to solve them.

 We are here going to summarize final mathematical conclusions reported in Ref.~\cite{Mess-Nak} to which the reader should refer
    for any mathematical detail concerning this procedure. With reference to the problem under scrutiny, the entries $a(t)$ and $b(t)$ of the unitary operator $U(t)$ solution
    of \eqref{16_1} may be cast as follows
  \begin{eqnarray}\label{16_8}
  a(t)&=&\frac{1}{\sqrt{1+|X|^2}}e^{-i\int_{0}^{t}\Omega(t')dt'-i\int_{0}^{t}\frac{\text{Im}[\dot{X}X^{*}]}{1+|X|^2}dt'},
\end{eqnarray}
\begin{eqnarray}\label{16_7}
b(t)=-ia(t)X(t),
\end{eqnarray}
where $X(t)$ is an arbitrary complex function of $t$, with the initial condition $X(0)=0$, prescribing the link between $\Omega(t)$ and $\omega(t)$ through the equation
\begin{eqnarray}\label{16_44}\omega(t)=a^2(t)\dot{X}(t)\end{eqnarray}
in order to make the corresponding generalized Rabi problem exactly solvable.
In other words, once fixed the time dependence of the longitudinal component $\Omega(t)$ of the magnetic field of experimental interest in $H$ as well as $X(t)$
at will, equations \eqref{16_42} immediately provide the time evolution of a generalized Rabi system wherein the transversal components $\omega_x(t)=\text{Re}[\omega(t)]$, $\omega_y(t)=-\text{Im}[\omega(t)]$ of the applied magnetic field must be engineered in accordance with \eqref{16_44}.
\par In the next section we are going to show how this "knowhow"  may be exploited to solve a special class of Riccati equations.

%is introduced.

%where the initial condition $U(0)=I$ is given.

%Using it, the first equation in \eqref{16_32} can be rewritten as

%Since it is a challenging problem to find the general solution of the latter differential equations, the authors limited their searching by the exactly solvable cases.
%The integro-differential equation \eqref{16_4} can be rewritten as a differential as
% \begin{eqnarray*}\dot{a}=-\left(\frac{i}{\hbar}\Omega+\frac{\dot{X}X^{*}}{\hbar^2+|X|^2}\right)a.
%\end{eqnarray*}
%Its solution in terms of the function $X$ is the following

%The latter approach is a mathematical trick, but it helps to find the explicitly solvable Hamiltonian problems. All the relations are exact and no approximation is required. Chosen some function $X(t)$, that gives the connection \eqref{16_4_1} and an arbitrary longitudinal field $\Omega(t)$ the transverse field $\omega(t)$ is given by \eqref{16_7} and the unitary evolution operator is given by the matrix with the matrix elements $a$ and $b$ defined by \eqref{16_8} and \eqref{16_7}. Note, that since the most general form of the Hamiltonian is considered, the MN approach can be applied to any two-level system scenario.

\section{Towards the new Riccati Equation}\label{sec:3}
We start from the coupled equations \eqref{16_42} defining the Cauchy problem whose resolution provides the exact quantum dynamics of any two-state
physical scenario describable by the Hamiltonian given by equation \eqref{16_45}.
Let us introduce a new complex-valued function $u(t)$ related to $a(t)$ and $b(t)$ as follows
\begin{eqnarray}\label{16_25}
u(t)a^{*}(t)=b(t),%,\quad u_2(t)=\ln{a^{*}(t)},\quad a^{*}(t)u_3(t)=-b^{*}(t),
\end{eqnarray}
which implies $u(0)=0$. We wish to prove that if the pair of functions $a(t)$ and $b(t)$ solve the Cauchy problem \eqref{16_42}
then the correspondent function $u(t)$ satisfies a Riccati equation whose coefficients are expressible in terms of $\Omega(t)$ and $\omega(t)$.
%
%To this end we firstly derive Eq. \eqref{16_25}
%\begin{eqnarray}
%\dot{u}(t)a^{*}(t)+u(t)\dot{a}^{*}(t)=\dot{b}(t),
%\end{eqnarray}
It is easy to check that $u(t)$ satisfies the following differential equation
\begin{eqnarray}\label{16_23}
&&\dot{u}(t)=i\omega^{*}(t) u(t)^2-2i\Omega(t) u(t)-i\omega(t),%,\nonumber\\
%&&\dot{u}_2(t)=-\frac{\Omega(t)}{i\hbar}+\frac{\omega^{*}(t)}{i\hbar}u_1(t),\\\nonumber
%&&\dot{u}_3(t)=\frac{\omega^{*}(t)}{i\hbar}e^{-2u_2(t)}.
\end{eqnarray}
and the initial condition $u(0)=0$.
This equation, hereafter referred to as ``associated to the generalized Rabi problem'' under scrutiny, is a DRE having  the general form
%Hence $u_1^{*}(t)a(t)=-a^{*}(t)u_3(t)$ and from $aa^{*}+bb^{*}=1$ we can deduce that
% \begin{eqnarray*}e^{u_2(t)+u^{*}_2(t)}-u_1(t)e^{2u_2(t)}u_3(t)=1.
%\end{eqnarray*}
%Thus, the conditions on the new functions are
% \begin{eqnarray*}u_1(t)=-e^{u_2^{*}(t)-u_2(t)}u_3^{*},\quad |u_1(t)|^2=e^{-u_2^{*}(t)-u_2(t)}-1.
%\end{eqnarray*}
%Let us assume that $a^{*}(t)\neq 0$. We substitute \eqref{16_8} and \eqref{16_7} in \eqref{16_25} and rewrite the latter functions in terms of \eqref{16_4_1} as
 %\begin{eqnarray*}\label{Relations ui-ab}
% u_1(t)&=&\frac{b(t)}{a^{*}(t)}=\frac{1}{i\hbar}\frac{aX}{a^{*}},\quad u_2(t)=\ln{a^{*}(t)},\quad u_3(t)=-\frac{b^{*}(t)}{a^{*}(t)}=\frac{1}{i\hbar}X^{*}.
%\end{eqnarray*}
%From \eqref{16_25}
% \begin{eqnarray*}a^{*}(t)=e^{u_2(t)},\quad b^{*}(t)=-e^{u_2(t)}u_3(t),\quad b(t)=e^{u_2(t)}u_1(t)
%\end{eqnarray*}
%hold. Substituting the latter expressions in \eqref{16_32}, we can obtain the following system of the differential equations on $u_1(t)$, $u_2(t)$ and $u_3(t)$
%\begin{eqnarray}\label{16_23}&&\dot{u}_1(t)=-\frac{\omega^{*}(t)}{i\hbar}u_1(t)^2+2\frac{\Omega(t)}{i\hbar}u_1(t)+\frac{\omega(t)}{i\hbar}%,\nonumber\\
%&&\dot{u}_2(t)=-\frac{\Omega(t)}{i\hbar}+\frac{\omega^{*}(t)}{i\hbar}u_1(t),\\\nonumber
%&&\dot{u}_3(t)=\frac{\omega^{*}(t)}{i\hbar}e^{-2u_2(t)}.
%\end{eqnarray}
%One can see, that the first equation in \eqref{16_23} is the Riccati equation, i.e.
\begin{eqnarray}\label{16_18}\dot{y}(t)=f(t)y^2(t)+g(t)y(t)+f^*(t)
\end{eqnarray}
with $f(t)$ and $g(t)$, in general, complex-valued functions.
To the best of the present authors' knowledge, no tool is known in literature to determine a particular solution of equation \eqref{16_18} and then its general integral whatever $f$ and $g$ are.
Therefore, such an equation introduces a special class of DREs, which we call `new' in the sense that it is not included in the list of solved Riccati equations given in \cite{Zwillinger,Polyanin}.

First of all, it is important to point out that if we consider a physical scenario (assigned $\Omega$ and $\omega$) for which the solution of the related Cauchy problem \eqref{16_42} is known, then we may write easily the solution of the associated DRE through Eq. \eqref{16_25}.
For example, in \cite{GMN} the authors show explicitly the solution of the problem considering the following two scenarios with constant $|\omega|$ ($\omega(t)=|\omega|e^{i\phi_\omega(t)}$)
\begin{subequations}
\begin{align}
& \Omega({t}) = \dfrac{2 |\omega|}{\cosh(2|\omega|t)} - {\dot{\phi}_\omega({t}) \over 2}, \\
& \Omega({t}) = {|\omega| \over 2} \biggl[ {3 \over \cosh(|\omega|t)} - \cosh (|\omega|t) \biggr] - {\dot{\phi}_\omega({t}) \over 2}.
\end{align}
\end{subequations}
The related solutions satisfying the Cauchy problem \eqref{16_42} read respectively ($a=|a|e^{i\phi_a}, b=|b|e^{i\phi_b}$)
\begin{subequations}
\begin{align}
&|a(t)| = \sqrt{\dfrac{ \cosh(2|\omega|t) + 1 }{2 \cosh(2|\omega|t)}}, \quad \phi_{a}(t) = \dfrac{\phi_\omega(t)}{2} - \tan^{-1} \Bigl[ \tanh \Bigl( {|\omega|t} \Bigr) \Bigr] - {|\omega|t}, \\
&|b(t)| = \sqrt{\dfrac{ \cosh(2|\omega|t) - 1 }{2 \cosh(2|\omega|t)}}, \quad \phi_{b}(t) = \phi_{a}(t) + 2|\omega|t - {\pi \over 2},
\end{align}
\end{subequations}
and
\begin{subequations}
\begin{align}
&|a(t)| = \dfrac{1}{\cosh(|\omega|t)}, \qquad \phi_{a}(t) = \dfrac{\phi_\omega(t)}{2} - \tan^{-1} \Bigl[ \tanh \Bigl( {|\omega|t\over 2} \Bigr) \Bigr] - {\sinh(|\omega|t) \over 2}, \\
&|b(t)| = \tanh(|\omega|t), \qquad \phi_{b}(t) = \phi_{a}(t) + \sinh(|\omega|t)- {\pi \over 2}.
\end{align}
\end{subequations}
In these cases, it can be verified that the associated DREs and their particular [$u(0)=0$] integrals turn out to be respectively
\begin{subequations}
\begin{align}
&\dot{u}=i \Bigl[ \omega^* u^2+\Bigl( \dot{\phi}_\omega(t) - {4 |\omega| \over \cosh(2|\omega|t)} \Bigr) u - \omega \Bigr], \\
&u(t)=\tanh(|\omega|t) \exp\Bigl\{ i \Bigl[\phi_\omega(t)-2 \tan^{-1}[\tanh(|\omega|t)]-\pi/2 \Bigr] \Bigr\} \label{special int case 1},
\end{align}
\end{subequations}
and
\begin{subequations}
\begin{align}
&\dot{u}=i \Bigl\{ \omega^* u^2+\Bigl[ {\dot{\phi}_\omega(t)} - |\omega| \Bigl( {3 \over \cosh(|\omega|t)} - \cosh (|\omega|t) \Bigr) \Bigr] u - \omega \Bigr\}, \\
&u(t)=\sinh(|\omega|t) \exp\Bigl\{ i \Bigl[\phi_\omega(t)-2 \tan^{-1}[\tanh(|\omega|t/2)]-\pi/2 \Bigr] \Bigr\} . \label{special int case 2}
\end{align}
\end{subequations}

We stress that in Eqs. \eqref{special int case 1} and \eqref{special int case 2} $|\omega|$ and $\phi_\omega(t)$ may be assigned at will.
We may better exploit the method reported in Ref. \cite{Mess-Nak} also to ``construct'' new DREs and at the same time their integrals.
Suppose, indeed, that the solution of a Cauchy problem expressed by the equations \eqref{16_42} has been found, in particular with the help of a smart auxiliary complex-valued function $X(t)$ in accordance with the recipe described in the previous section. Then, in view of equations \eqref{16_8} and \eqref{16_7}, we explicitly know both the expressions of $a(t)$ and $b(t)$ which, in turn, immediately implies that we know the particular solution $\bar{u}(t)$ [$\bar{u}(0)=0$] of the DRE associated to the corresponding general Rabi problem.
Such a solution may be written down as follows
\begin{equation}\label{16_47}
\bar{u}(t)=-{i} X(t) e^{-2i \bigl(\int_{0}^{t}{\Omega(t')}dt'+\int_{0}^{t}\frac{\text{Im}[\dot{X}X^{*}]}{1+|X|^2}dt'\bigr)},
\end{equation}
where $X(t)$ is the arbitrary function of $t$ introduced in Ref. \cite{Mess-Nak}. %to solve exactly the integral-differential equation \eqref{16_4} in $a$.
The knowledge of $\bar{u}(t)$ is enough to write down the general integral of equation \eqref{16_23} as follows \cite{Reid}
\begin{subequations}
\begin{align}
&u(t)=\bar{u}(t)+\Phi(t)\left(C-i\int_0^t \omega^{*}(t')\Phi(t')dt'\right)^{-1}, \\
&\Phi(t)=\exp\left\{2i\int_0^t \left[\omega^{*}(t')\bar{u}(t')-\Omega(t')\right]dt'\right\},
\end{align}
\end{subequations}
where \cite{Mess-Nak}
\begin{eqnarray}
&&\omega(t)=a^2(t)\dot{X}(t)=\frac{\dot{X}(t)}{1+|X(t)|^2}e^{-2i\int_{0}^{t}\Omega(t')dt'-2i\int_{0}^{t}\frac{\text{Im}[\dot{X}X^{*}]}{1+|X|^2}dt'}.\end{eqnarray}

Summing up: if the quantum dynamics of the generalized Rabi system, defined by $\Omega(t)$ and $\omega(t)$, is exactly solvable by means of the $X(t)$-based recipe, then a particular integral of the associated DRE may be cast in the form given in equation \eqref{16_47}. 
We point out that in $\bar{u}(t)$ $X(t)$ plays the role of a ``free parameter function'' since it may be assigned at will.
In addition $X(t)$ may be considered as a ``generator'' of the ``resolutive'' $\omega(t)$, which means that from the knowledge of $\Omega(t)$ and the choice of $X(t)$, the strategy given in \cite{Mess-Nak} provides the specific $\omega(t)$ generating an exactly solvable Rabi problem \eqref{16_42} and consequently a solvable associated DRE \eqref{16_23}.
We however emphasize that our procedure is truly successful when we are able to find  the explicit form of $\omega(t)$, that is when we are able to evaluate the explicit expression of $\phi_\omega(t)$.
Otherwise, this procedure cannot be effectively finalized, meaning that we do not have at our disposal the Riccati equation to be solved. 
Due to the wide variability of $X(t)$ leading to exactly solvable Rabi problems, we may claim that we are in condition to solve as well a wide class of DREs of the new type, that is, having the general structure given by \eqref{16_18}. 

Putting the question in general terms, if we make the correspondence $f(t)\leftrightarrow -i\omega(t)$, $g(t)\leftrightarrow -2i\Omega(t)$ between the coefficients of the associated Riccati equation and those appearing in the general equation \eqref{16_18} we may convert $\bar{u}(t)$ into the particular solution $\bar{y}(t)$ of \eqref{16_18}, namely
%\begin{eqnarray}\label{16_29}&&\bar{y}(t)=\frac{a(t)}{ a^{*}(t)}\int_{0}^{t}\frac{f(t')}{a^2(t')}dt',\quad \dot{a}(t)=-f(t)\int_{0}^{t}\frac{f^{*}(t')}{{a^{*}}^2(t')}dt'+\frac{g(t) a(t)}{2}
%\end{eqnarray}
%Using MN approach we may rewrite  it as
\begin{equation}\label{16_40}
\bar{y}(t)=-i X(t) e^{\int_{0}^{t}g(t')dt'-2i\int_{0}^{t}\frac{\text{Im}[\dot{X}X^{*}]}{1+|X|^2}dt'},
\end{equation}
which satisfies the condition $y(0)=0$ since $X(0)=0$.

This means that, once again with the help of Ref. \cite{Mess-Nak}, we are able to ``construct'' the function $f(t)\leftrightarrow -i\omega(t)$, by Eqs. \eqref{16_8} and \eqref{16_7}, in terms of the `free' function $g(t)\leftrightarrow -2i\Omega(t)$ and the arbitrary function $X(t)$, such that the new type of the Riccati equation may be exactly solved.
The particular solution, in this manner, may be written as prescribed by Eq. \eqref{16_40}.

In the following section we illustrate our procedure giving explicit examples of resolution of Riccati equations not appearing in the available specialized literature.

\section{Examples}\label{sec:4}

In this section we illustrate how, exploiting the general procedure developed in the previous section and considering specific forms for the function $X(t)$, we succeed in recovering new DREs as well as, at the same time, their exact solution.

\subsection{Example 1}
%\subsubsection{Example 1: Given Hamiltonian parameters}
Let us consider a complex function $X(t)$ parametrized by a real function $\phi(t)$ and a real constant parameter $c$ as
\begin{eqnarray}\label{16_31}
X(t)=c\sin[\phi(t)]e^{i\phi(t)},\quad c \in \mathbb{R},\quad  \phi(0)=0.
\end{eqnarray}
Following Ref. \cite{Mess-Nak} we choose the complex transverse magnetic field $\omega=|\omega|e^{i\phi_\omega}$ acting on the two-level system in the form (some time-dependences are left out for easier readings)
%&a=\frac{1}{\sqrt{1+c^2\sin^2{\phi}}}e^{-i\int_{0}^{t}\Omega dt'-i\int_{0}^{t}\frac{c^2\dot{\phi}\sin^2{\phi}}{1+c^2\sin^2{\phi}}dt'},\\
\begin{equation}\label{Special Relation}
\omega(t)=\frac{c\dot{\phi}}{1+c^2\sin^2{\phi}}e^{-2i\int_{0}^{t}\Omega dt'+2i\int_{0}^{t}\frac{\dot{\phi}}{1+c^2\sin^2{\phi}}dt'}.
%&|X|^2=c^2\sin^2{\phi},\quad \dot{X}=ce^{2i\phi}\dot{\phi},\quad \dot{X}^{*}=ce^{-2i\phi}\dot{\phi},
\end{equation}
getting, in addition the following relation between the complex  $\omega$ and $\Omega$
\begin{eqnarray}\label{16_30}
&&\frac{|\omega(t)|}{c}=\Omega(t)+\frac{{\dot{\phi}}_{\omega}}{2}.
\end{eqnarray}
To construct explicit expressions of $a(t)$ and $b(t)$ getting rid of the parameter function $X(t)$ in Eqs. \eqref{16_8} and \eqref{16_7}, following Ref. \citep{Mess-Nak}, the arbitrary parameter function $\phi(t)$ is written in terms of $|\omega(t)|$ as follows
\begin{eqnarray}
&&\tan{\phi}=\frac{1}{\sqrt{1+c^2}}\tan{\Phi(t)},\quad \Phi(t)=\frac{\sqrt{1+c^2}}{c}\int_{0}^{t}|\omega(t')|dt'.
\end{eqnarray}
This immediately leads to expressions of $a(t)$ and $b(t)$ only in terms of the Hamiltonian parameters $\omega(t)$ and $\Omega(t)$ as follows
\begin{subequations}
\begin{align}
&a(t)=\sqrt{\frac{1+c^2\cos^2{\Phi}}{1+c^2}}e^{-i\int_{0}^{t}\Omega(t')dt'+\frac{i}{c}\int_{0}^{t}|\omega(t')|dt'-i\tan^{-1}{\left(\frac{1}{\sqrt{1+c^2}}\tan{\Phi}\right)}},\\
&b(t)=-i\frac{c\sin{\Phi}}{\sqrt{1+c^2}}e^{-i\int_{0}^{t}\Omega(t')dt'+\frac{i}{c}\int_{0}^{t}|\omega(t')|dt'}.
\end{align}
\end{subequations}
Hence, using \eqref{16_25}, we may easily recover the particular solution $\bar{u}(t)$ of Eq. \eqref{16_23} satisfying the initial condition $\bar{u}(0)=0$, namely
\begin{eqnarray}\label{16_21}
&&\bar{u}(t)=\frac{-ic\sin{\Phi}}{\sqrt{1+c^2\cos^2{\Phi}}}e^{-2i\int_{0}^{t}\Omega(t')dt'+\frac{2i}{c}\int_{0}^{t}|\omega(t')| dt'}
e^{-i\tan^{-1}{\left(\frac{1}{\sqrt{1+c^2}}\tan{\Phi}\right)}}.
\end{eqnarray}

In more general terms, we claim that the Riccati equation
\begin{align}\label{16_34}
\dot{y}(t)=f^{*}(t)y(t)^2+g(t)y(t)+f(t),
\end{align}
has the following particular solution [$y(0)=0$]
\begin{eqnarray}\label{16_22}
&&y(t)=\frac{-ic\sin{\left(\frac{\sqrt{1+c^2}}{c}\int_{0}^{t}|f(t')|dt'\right)}}{\sqrt{1+c^2\cos^2{\left(\frac{\sqrt{1+c^2}}{c}\int_{0}^{t}|f(t')|dt'\right)}}}
e^{-i{\phi}_{f}(t)
-i\tan^{-1}{\left[\frac{1}{\sqrt{1+c^2}}\tan{\left(\frac{\sqrt{1+c^2}}{c}\int_{0}^{t}|f(t')|dt'\right)}\right]}},
\end{eqnarray}
if the following link between $f \equiv |f|e^{i\phi_f}$ and $g$ is postulated
\begin{equation}
\frac{|f(t)|}{c}=\frac{ig(t)}{2}+\frac{{\dot{\phi}_{f}(t)}}{2},\quad \phi_{f}(0)=0. \label{Special relation f-g}
\end{equation}

This equation gives the key to identify a class of exactly solvable DREs.
Indeed, the general relation \eqref{16_30} between the Hamiltonian parameters and then between the DRE coefficients in \eqref{Special relation f-g} is a sufficient condition ensuring that the associated DRE \eqref{16_34} can be solved analytically, its explicit exact particular solution being expressed by Eq. \eqref{16_22}.

\subsection{Example 2}
In this subsection we consider the most general form for the function $X(t)$, that is, a general complex-valued function, namely
\begin{equation}
X(t)=A(t)e^{i\phi(t)},
\end{equation}
with $A(t)$ and $\phi(t)$ real-valued functions, of course.
The sufficient condition between the parameters $\Omega(t)$ and $\omega(t)$ in \eqref{16_45}, so that the related Cauchy problem \eqref{16_1} is analytically exactly solvable, reads \cite{Mess-Nak} (from now on the time-dependence of the variables in the formulas is left out)
\begin{equation}\label{Gen Rel Om-om}
\Omega={1\over2}(\dot\Theta-\dot\phi_\omega)+|\omega|\sin\Theta\cot\Bigl[2\int_0^t|\omega|\cos\Theta dt'\Bigr],
\end{equation}
with $\tan\Theta={A \over \dot{A}} \dot{\phi}$ and $\Theta(0)=0$.
In this case the solutions for the two parameters $a$ and $b$ satisfying the system \eqref{16_42} may be cast as follows 
\begin{subequations}
\begin{align}
&|a|=\cos\left[\int_0^t|\omega|\cos\Theta dt'\right], \qquad \phi_a={\phi_\omega \over 2}-{\Theta\over 2}-\mathcal{R}, \\
&|b|=\sin\left[\int_0^t|\omega|\cos\Theta dt'\right], \qquad \phi_b={\phi_\omega \over 2}-{\Theta\over 2}+\mathcal{R}-{\pi \over 2},
\end{align}
\end{subequations}
where we put
\begin{equation}
\mathcal{R}=2\int_0^t{|\omega|\sin\Theta\over\sin\bigl[2\int_0^{t'}|\omega|\cos\Theta dt''\bigr]} dt'.
\end{equation}

In this manner we may write the solution of the DRE \eqref{16_23} with $\Omega(t)$ and $\omega(t)$ linked by Eq. \eqref{Gen Rel Om-om} as (see Eq. \eqref{16_25})
\begin{equation}\label{Solution RE more Gen}
u(t)={b\over a^*}={|b|\over |a|}e^{i(\phi_b+\phi_a)}=\tan\left[\int_0^t|\omega|\cos\Theta dt'\right]\exp\left\{ i \left[\phi_\omega(t) - \Theta - {\pi \over 2}\right] \right\}.
\end{equation}
We note that in this case the solution depends explicitly on the arbitrary function $\Theta(t)$, meaning that choosing at our will both the parameter $\omega(t)$ and the function $\Theta(t)$, and moreover ``constructing'' the other parameter $\Omega(t)$ as prescribed in Eq. \eqref{Gen Rel Om-om}, Eq. \eqref{Solution RE more Gen} gives us the solution of the associated DRE in Eq. \eqref{16_23}.
To make clear this procedure and to show the applicability and the usefulness of this method, three exemplary cases are reported in the following subsections.

\subsubsection{Specific Examples}
If we choose (given a particular $\omega(t)$)
\begin{equation}
\Theta=\int_0^t |\omega| dt'
\end{equation}
we get
\begin{subequations}
\begin{align}
&\Omega={1\over2}(|\omega|-\dot\phi_\omega)+|\omega|\sin\Bigl( \int_0^t |\omega| dt' \Bigr)\cot\Bigl[2\sin\Bigl( \int_0^t |\omega| dt' \Bigr)\Bigr], \\
&|a|=\biggl|\cos\Bigl[\sin\Bigl( \int_0^t |\omega| dt' \Bigr)\Bigr] \biggr|, \quad |b|=\biggl| \sin\Bigl[\sin\Bigl( \int_0^t |\omega| dt' \Bigr)\Bigr] \biggr|, \\
&\phi_a+\phi_b=\phi_\omega-\int_0^t |\omega| dt' - \pi/2
\end{align}
\end{subequations}
and it is possible to verify that the solution of the corresponding DRE \eqref{16_23} is
\begin{equation}
\bar{u}(t)=\tan\Bigl[\sin\Bigl( \int_0^t |\omega| dt' \Bigr)\Bigr] \exp\biggl\{ i \Bigl( \phi_\omega-\int_0^t |\omega| dt'-\pi/2 \Bigr) \biggr\}
\end{equation}
according to Eq. \eqref{Solution RE more Gen}.

If we now consider a transversal field with a constant magnitude $|\omega(t)|=|\omega(0)|\equiv |\omega|$ and choose
\begin{equation}
\Theta = 2 \tan^{-1}\left({2|\omega|t \over \sqrt{2+4(|\omega|t)^2}}\right),
\end{equation}
we get
\begin{subequations} \label{Omega a and b case 1}
\begin{align}
&\Omega = 4|\omega| {1+(|\omega|t)^2 \over [1+4(|\omega|t)^2] \sqrt{2+4(|\omega|t)^2}} -{\dot{\phi}_\omega(t) \over 2} , \\
&|a| = \sqrt{{\sqrt{1+4(|\omega|t)^2} + 1 \over 2 \sqrt{1+4(|\omega|t)^2}}}, \quad  |b| = \sqrt{{\sqrt{1+4(|\omega|t)^2} - 1 \over 2 \sqrt{1+4(|\omega|t)^2}}}, \\
&\phi_a+\phi_b = {\dot{\phi}_\omega} - 2 \tan^{-1}\left({2|\omega|t \over \sqrt{2+4(|\omega|t)^2}}\right) - {\pi \over 2}.
\end{align}
\end{subequations}
In this manner we have the following DRE
\begin{equation}
\dot{u}=i \left[ \omega^* u^2+\left( \dot{\phi}_\omega - 8|\omega| {1+(|\omega|t)^2 \over [1+4(|\omega|t)^2] \sqrt{2+4(|\omega|t)^2}} \right) u - \omega \right]
\end{equation}
and it is possible to verify that its particular solution [$u(0)=0$] reads
\begin{equation}
\bar{u}(t)=\sqrt{{\sqrt{1+4(|\omega|t)^2} - 1 \over \sqrt{1+4(|\omega|t)^2} + 1}} \exp\left\{ i \left[\phi_\omega(t)-2 \tan^{-1}\left({2|\omega|t \over \sqrt{2+4(|\omega|t)^2}}\right)-\pi/2 \right] \right\},
\end{equation}
according to Eq. \eqref{Solution RE more Gen}.

It is interesting to note that if we put
\begin{equation}
\Theta = 2 \tan^{-1}\left({|\omega|t \over \sqrt{2+(|\omega|t)^2}}\right),
\end{equation}
we get
\begin{subequations}
\begin{align}
&\Omega= |\omega| \left[ {2+[1-(|\omega|t)^2][2+(|\omega|t)^2] \over 2[1+(|\omega|t)^2]\sqrt{2+(|\omega|t)^2}}\right]-{\dot{\phi}_\omega(t) \over 2}, \\
&|a| = {1 \over \sqrt{1+(|\omega|t)^2}}, \qquad |b| = {|\omega|t \over \sqrt{1+(|\omega|t)^2}} = |a(t)||\omega|t, \\
&\phi_a+\phi_b = {\dot{\phi}_\omega} - 2 \tan^{-1}\left({|\omega|t \over \sqrt{2+(|\omega|t)^2}}\right) - {\pi \over 2}.
\end{align}
\end{subequations}
In this case the associated DRE reads
\begin{eqnarray}
\dot{u}=i \left\{ \omega^* u^2+\left[ {\dot{\phi}_\omega(t)} - 2|\omega| \left( {2+[1-(|\omega|t)^2][2+(|\omega|t)^2] \over 2[1+(|\omega|t)^2]\sqrt{2+(|\omega|t)^2}}\right) \right] u - \omega \right\}
\end{eqnarray}
whose solution results
\begin{equation}
\bar{u}(t)=|\omega| t \exp\left\{ i \left[\phi_\omega(t)-2 \tan^{-1}\left({|\omega|t \over \sqrt{2+(|\omega|t)^2}}\right)-\pi/2 \right] \right\}.
\end{equation}
We note how a slight change in the arbitrary choice of the function $\Theta$ leads to a very different associated DRE.

%We remark finally that these last three cases offer three physical scenarios, that is expressions for the time-dependence of the Hamiltonian parameters, whose related Cauchy-problems \eqref{16_22} may be solved analytically.
%Therefore, they are clear examples of the interplay existing between solvable DREs and the analytical derivation of SU(2)-symmetric dynamics, that is how it were possible through our approach to recover exactly solvable SU(2)-symmetric physical problems by the resolution of specific DREs and vice versa.

\section{Conclusive Remarks}\label{sec:5}

In the interplay between physics and mathematics, the latter, more often, provides systematic or tricky tools to solve basic or derived equations of crucial interest in making predictions concerning the behaviour of a specific physical problem.
Under this point of view the DRE occupies a special place due to its unifying character \citep{Schuch}, as shown by Euler (\citep{Bittanti} and references therein).
To appreciate this peculiar property it is enough, for example, to mention that the unidimensional Schr\"odinger equation of a single particle may be always traced back to a special DRE \citep{Haley} or to consider the role that such a nonlinear differential equation plays when the concept of supersymmetry is applied to quantum mechanics \citep{Filho,Cooper}.

In this paper we provide a remarkable example of how a resolutive strategy envisioned for a class of physical problems may be smartly exploited and adapted for constructing the general integral of a special class of DRE, for the first time.
This kind of application to the best of the present authors' knowledge is original and leads to a protocol to generate exactly solvable DREs whose coefficients are related through a parametric function $X(t)$ or $\Theta(t)$ which may be prescribed at will.
We illustrate its applicability considering several examples of non-trivial DREs for which it turns out to be easy to provide explicit solutions. 

% the roles of physics and mathematics in their interplay providing a remarkable example of resolution
%The main goal reached in this paper is the individuation and the exact resolution of special DREs.
%Our approach is based on the knowledge of a recently reported strategy \citep{Mess-Nak} for generating the exact quantum dynamics of a time-dependent bidimensional SU(2) problem.
%This point of view, as far as we know, is original and leads to our main results.
%%First of all, the coupled differential equations governing the time-dependence of the two entries of the unitary evolution operator $U$, $a$ and $b$, relative to a generic bidimensional SU(2) problem, are converted into a DRE of a new time-dependent variable $u$ smartly related to the two entries of $U$.
%%The second step furnishes indeed an exact solution of such a DRE, as soon as we are able to solve the associated SU(2) problem under scrutiny.
%
%In this paper indeed we have shown how to exploit the strategy reported in Ref. \citep{Mess-Nak} to individuate and solve SU(2) bidimensional problems, in order to find a particular solution of the associated DRE and then its general integral.
%This protocol thus represents a remarkable example of interplay between physics and mathematics wherein the strategy to solve a mathematical problem stems from that used to solve a physical problem.
%In this paper hence we provide and illustrate, with several applications, a protocol to generate exactly solvable DREs whose coefficients are related through a parametric function which may be prescribed at will.
We emphasize that our approach is rather different from that reported in Ref. \citep{Harko} where the solutions of some DREs are obtained prescribing specific links between the coefficients appearing in the nonlinear differential equation.
In addition we observe that the point of view reported in this paper may be reversed, meaning that, in principle, the knowledge of the general integral of a DRE could be exploited to individuate and solve SU(2) quantum dynamical problems.

In conclusion we underline that the results reported in this paper are entirely based on ideas and method given in Ref. \cite{Mess-Nak}, which, besides having been already applied to treat time-dependent spin Hamiltonian models beyond the relatively simpler $s=1/2$ case \cite{GMN,GMIV,GBNM}, has even the merit of offering tools to solve a wide class of special DREs not considered in the specialized literature so far.

\section*{Acknowledgements}
%The study in section \ref{sec:3} by
A. Messina and R. Grimaudo warmly thank Prof. A. Klimov and Prof. M. Kus for interesting and stimulating discussions about the argument.
H. N. was supported by the Waseda University Grant for Special Research Projects (No. 2016B-173).
The study in sections 2 and 3 by Markovich L.A. was supported by the Russian Science Foundation grant (14-50-00150).
R. G. was partially supported by research funds in memory of Francesca Palumbo.

%% References


\bibliographystyle{elsarticle-num}
% \bibliographystyle{elsarticle-harv}
% \bibliographystyle{elsarticle-num-names}
% \bibliographystyle{model1a-num-names}
% \bibliographystyle{model1b-num-names}
% \bibliographystyle{model1c-num-names}
% \bibliographystyle{model2-names}
% \bibliographystyle{model3a-num-names}
% \bibliographystyle{model3-num-names}
% \bibliographystyle{model4-names}
% \bibliographystyle{model5-names}
% \bibliographystyle{model6-num-names}

%\nocite{*}
\bibliography{Bibliography} 

\begin{thebibliography}{10}
\expandafter\ifx\csname url\endcsname\relax
  \def\url#1{\texttt{#1}}\fi
\expandafter\ifx\csname urlprefix\endcsname\relax\def\urlprefix{URL }\fi
\expandafter\ifx\csname href\endcsname\relax
  \def\href#1#2{#2} \def\path#1{#1}\fi

\bibitem{Reid}
W.~T. Reid, Riccati differential equations, Elsevier, 1972.

\bibitem{Fraga}
S.~Fraga, J.~G. de~la Vega, The Schr{\"o}dinger and Riccati Equations,
  Springer-Verlag, 1998.

\bibitem{Laub}
S.~Bittanti, A.~J. Laub, J.~C. Willems, Invariant subspace methods for the
  numerical solution of riccati equations, in: The Riccati Equation, Springer,
  1991, pp. 163--196.

\bibitem{Schuch}
D.~Schuch, in: Journal of Physics: Conference Series, Vol. \textbf{504}, IOP
  Publishing, 2014, p. 012005.

\bibitem{Schuch1}
D.~Schuch, in: Journal of Physics: Conference Series, Vol. \textbf{538}, IOP
  Publishing, 2014, p. 012019.

\bibitem{Rau}
A.~Rau, Physical Review A \textbf{61}~(3) (2000) 032301.

\bibitem{Rau1}
A.~Rau, Physical review letters \textbf{81}~(22) (1998) 4785.

\bibitem{Harko}
T.~Harko, F.~S. Lobo, M.~Mak, Universal Journal of Applied Mathematics
  \textbf{2}~(2) (2014) 109--118.

\bibitem{Liverts}
E.~Liverts, V.~Mandelzweig, Physica Scripta \textbf{77}~(2) (2008) 025003.

\bibitem{Unal}
N.~{\"U}nal, Annals of Physics \textbf{327}~(9) (2012) 2177--2183.

\bibitem{Gardas}
B.~Gardas, \textit{Riccati equation in studies of spin-boson systems}, ph.d.
  thesis.

\bibitem{Bastami}
A.~Al~Bastami, M.~R. Belic, N.~Z. Petrovic, Electronic Journal of Differential
  Equations \textbf{2010}~(66) (2010) 1--10.

\bibitem{Suazo}
E.~Suazo, S.~K. Suslov, J.~M. Vega-Guzm{\'a}n, Mathematics \textbf{2}~(2)
  (2014) 96--118.

\bibitem{Kus}
S.~Charzy{\'n}ski, M.~Ku{\'s}, Journal of Physics A: Mathematical and
  Theoretical \textbf{46}~(26) (2013) 265208.

\bibitem{Kus1}
J.~Gutt, S.~Charzy{\'n}ski, M.~Ku{\'s}, Differential Geometry and its
  Applications \textbf{42} (2015) 37--43.

\bibitem{Kus2}
S.~Charzy{\'n}ski, M.~Ku{\'s}, Journal of Differential Equations
  \textbf{259}~(4) (2015) 1542--1559.

\bibitem{Bougoffa}
L.~Bougoffa, The Scientific World Journal 2014.

\bibitem{Busawon}
K.~Busawon, P.~Johnson, in: Proceedings of the 8th WSEAS International
  Conference on Applied Mathematics, Citeseer, 2005, pp. 334--338.

\bibitem{Messina}
R.~Messina, M.~Jivulescu, A.~Messina, A.~Napoli, Mathematical Methods in the
  Applied Sciences \textbf{30}~(16) (2007) 2055--2064.

\bibitem{Mukherjee}
S.~Mukherjee, B.~Roy, International Journal of Nonlinear Science
  \textbf{14}~(2) (2012) 251--256.

\bibitem{jivulescu}
M.~Jivulescu, A.~Napoli, A.~Messina, Reports on Mathematical Physics
  \textbf{62}~(3) (2008) 369--387.

\bibitem{Panayotounakos}
D.~E. Panayotounakos, T.~I. Zarmpoutis, P.~Sotiropoulos, A.~G. Kostogiannis,
  Annals of the University of Craiova-Mathematics and Computer Science Series
  \textbf{39}~(2) (2012) 211--225.

\bibitem{Tesavrova}
Z.~Tesa{\v{r}}ov{\'a}, Archivum Mathematicum \textbf{18}~(3) (1982) 133--143.

\bibitem{Yamaleev}
R.~M. Yamaleev, Indian Journal of Pure and Applied Mathematics \textbf{45}~(2)
  (2014) 165--184.

\bibitem{Yuzbasi}
{\c{S}}.~Y{\"u}zba{\c{s}}i, M.~Kara{\c{c}}ayir, in: Journal of Physics:
  Conference Series, Vol. \textbf{766}, IOP Publishing, 2016, p. 012036.

\bibitem{Kengne}
E.~Kengne, F.~B. Hamouda, A.~Lakhssassi, Applied Mathematics \textbf{4}~(10)
  (2013) 1471.

\bibitem{Navickas}
Z.~Navickas, R.~Vilkas, T.~Telksnys, M.~Ragulskis, Journal of biological
  dynamics \textbf{10}~(1) (2016) 297--313.

\bibitem{Bittanti}
S.~Bittanti, in: Decision and Control, 1996., Proceedings of the 35th IEEE
  Conference on, Vol. \textbf{2}, IEEE, 1996, pp. 1599--1604.

\bibitem{Polyanin}
V.~F. Zaitsev, A.~D. Polyanin, Handbook of exact solutions for ordinary
  differential equations, CRC press, 2002.

\bibitem{Zwillinger}
D.~Zwillinger, Handbook of differential equations, Vol.~1, Gulf Professional
  Publishing, 1998.

\bibitem{Hasegawa}
K.~Hasegawa, The riccati equation and its applications in physics, Ph.D.
  thesis, Reed College (2001).

\bibitem{Klimov}
A.~B. Klimov, S.~M. Chumakov, John Wiley \& Sons, 2009.

\bibitem{Marmo}
J.~F. Cari{\~n}ena, A.~Ibort, G.~Marmo, G.~Morandi, Springer, 2015.

\bibitem{Haley}
S.~B. Haley, American Journal of Physics \textbf{65}~(3) (1997) 237--243.

\bibitem{Ronveaux}
A.~Ronveaux, American Journal of Physics \textbf{40}~(6) (1972) 888--892.

\bibitem{Ronveaux1}
A.~Ronveaux, American Journal of Physics \textbf{40}~(6) (1972) 892--896.

\bibitem{Santiago}
R.~Cruz-Santiago, J.~L{\'o}pez-Bonilla, J.~Morales, Himalayan Physics
  \textbf{4} (2013) 32--33.

\bibitem{Kryachko}
E.~S. Kryachko, Collection of Czechoslovak chemical communications
  \textbf{70}~(7) (2005) 941--950.

\bibitem{Sidharth}
B.~Sidharth, B.~Lakshmi, arXiv preprint physics/0305083.

\bibitem{Bougouffa}
S.~Bougouffa, S.~Al-Awfi, The European Physical Journal-Special Topics
  \textbf{160}~(1) (2008) 43--50.

\bibitem{Filho}
E.~Drigo~Filho, R.~M. Ricotta, Physics Letters A \textbf{269}~(5) (2000)
  269--276.

\bibitem{Mess-Nak}
A.~Messina, H.~Nakazato, Journal of Physics A: Mathematical and Theoretical
  \textbf{47}~(44) (2014) 445302.

\bibitem{Kuna}
M.~Kuna, J.~Naudts, Reports on Mathematical Physics \textbf{65}~(1) (2010)
  77--108.

\bibitem{Bagrov}
V.~G. Bagrov, D.~M. Gitman, M.~C. Baldiotti, A.~Levin, Annalen der Physik
  \textbf{14}~(11-12) (2005) 764--789.

\bibitem{DasSarma}
E.~Barnes, S.~D. Sarma, Physical review letters \textbf{109}~(6) (2012) 060401.

\bibitem{GMN}
R.~Grimaudo, A.~Messina, H.~Nakazato, Physical Review A \textbf{94}~(2) (2016)
  022108.

\bibitem{Cooper}
F.~Cooper, A.~Khare, U.~Sukhatme, Supersymmetry in quantum mechanics, World
  Scientific, 2001.

\bibitem{GMIV}
R.~Grimaudo, A.~Messina, P.~A. Ivanov, N.~V. Vitanov, Journal of Physics A:
  Mathematical and Theoretical \textbf{50}~(17) (2017) 175301.

\bibitem{GBNM}
R.~Grimaudo, Y.~Belousov, H.~Nakazato, A.~Messina, Quantum dynamics of two
  coupled spins under controllable and fluctuating magnetic fields, arXiv
  preprint arXiv:1703.07673.

\end{thebibliography}

\end{document}